\documentclass[journal]{IEEEtran}
\usepackage{multirow}
\usepackage{xcolor}
\usepackage{caption}
\usepackage{float}
\usepackage[colorlinks=TRUE]{hyperref}
\usepackage{graphicx}
\usepackage{amsmath}

\ifCLASSINFOpdf
\else
\fi
\begin{document}
%
\title{Stochastic Block Models with Multiple Continuous Attributes}
%
%
%

\author{
    \IEEEauthorblockN{Natalie Stanley\IEEEauthorrefmark{1}\IEEEauthorrefmark{5}, Thomas Bonacci \IEEEauthorrefmark{2}, Roland Kwitt \IEEEauthorrefmark{3}, Marc Niethammer\IEEEauthorrefmark{4}, Peter J. Mucha\IEEEauthorrefmark{5}} \\
    \IEEEauthorblockA{\IEEEauthorrefmark{1}Curriculum in Bioinformatics and Computational Biology, The University of North Carolina at Chapel Hill} \\
   \IEEEauthorblockA{\IEEEauthorrefmark{2}
    Department of Pharmacology \& Lineberger Comprehensive Cancer Center, The University of North Carolina at Chapel Hill} \\
     \IEEEauthorblockA{\IEEEauthorrefmark{3}
    Department of Computer Science, University of Salzburg}\\
     \IEEEauthorblockA{\IEEEauthorrefmark{4}
    Department of Computer Science \& Biomedical Research Imaging Center, The University of North Carolina at Chapel Hill}\\
    \IEEEauthorblockA{\IEEEauthorrefmark{5}
    Carolina Center for Interdisciplinary Applied Mathematics, Department of Mathematics, University of North Carolina-Chapel Hill
    \\ \IEEEauthorrefmark{1} stanleyn@email.unc.edu}
    
    }

\maketitle

\begin{abstract}
The stochastic block model (SBM) is a probabilistic model for community structure in networks. Typically, only the adjacency matrix is used to perform SBM parameter inference. In this paper, we consider circumstances in which nodes have an associated vector of continuous attributes that are also used to learn the node-to-community assignments and corresponding SBM parameters. While this assumption is not realistic for every application, our model assumes that the attributes associated with the nodes in a network's community can be described by a common multivariate Gaussian model. In this augmented, attributed SBM, the objective is to simultaneously learn the SBM connectivity probabilities with the multivariate Gaussian parameters describing each community. While there are recent examples in the literature that combine connectivity and attribute information to inform community detection, our model is the first augmented stochastic block model to handle multiple continuous attributes. This provides the flexibility in biological data to, for example, augment connectivity information with continuous measurements from multiple experimental modalities. Because the lack of labeled network data often makes community detection results difficult to validate, we highlight the usefulness of our model for two network prediction tasks: link prediction and collaborative filtering. As a result of fitting this attributed stochastic block model, one can predict the attribute vector or connectivity patterns for a new node in the event of the complementary source of information (connectivity or attributes, respectively). We also highlight two biological examples where the attributed stochastic block model provides satisfactory performance in the link prediction and collaborative filtering tasks.
\end{abstract}

\begin{IEEEkeywords}
Stochastic Block Model, Networks, Community Detection, Attributes
\end{IEEEkeywords}

\section{Introduction}

Uncovering patterns in network data is a common pursuit across a range of fields, such as in biology \cite{dan}, medicine \cite{agha,cancer} and computational social science \cite{socialnetwork}. A powerful way to analyze mesoscale structural organization within a network is with community structure \cite{muchacommunity,jurecommunity,shaiCC,fortu1,fortu2}. In this pursuit, the objective is to identify cohesive groups of nodes with relatively high density of within-group connections and fewer between-group connections. Numerous approaches exist to accomplish this task, but typically only the adjacency matrix encoding connectivity patterns is taken into account. In various applications, each node in a network is equipped with additional information (or particular attributes) that was not implicitly taken into account in the construction of the network. For example, in a protein interaction network, each protein could contain multiple experimental measurements or classifications. 

Significant attention has been given to the interplay between connectivity-based (or structural) community organization of the network and the attribute information of nodes within communities. Importantly, it is often unclear whether it is valid to assume that a \emph{structural} community should necessarily correlate with an attribute-based \emph{functional} community \cite{hric,peel2017ground,jureGroundTruth}. While such studies suggest that extreme caution should be taken in assuming a correlation between structural and functional communities, we limit our focus in the present work to the assumption that a node's connectivity and attribute patterns can be jointly modeled based on its community assignment. In other words, we seek to develop an approach to assign nodes to communities based jointly on both sources of information, such that a community is defined as a group of nodes with similar connectivity and attribute patterns. In doing so, our objectives are two-fold: first, we develop a probabilistic approach to jointly model connectivity and attributes; second, we wish to ensure that our model can handle multiple, continuous attributes.  

\subsection{Related work in attributed networks}
Recently, there have been numerous efforts to incorporate attribute information into the community detection problem \cite{cesna,clauset,ilouvain,hric,peel2017ground}. In describing our contribution, we distinguish between methods that descriptively obtain communities through optimization of a quality function and those that generatively capture communities through probabilistic models. Quality function based methods define a quantity of interest that an ideal partition would satisfy, while probabilistic methods identify communities through likelihood optimization and focus on the underlying statistical distribution for the observed network. A recent quality function-based method to handle multiple attributes is I-louvain \cite{ilouvain}. This method approaches the problem as an extension to the Louvain algorithm, which is the state-of-the-art scalable modularity quality function community detection method \cite{blondel}. The modularity-based approach to community detection defines a null model for community structure under the assumption that there is not substantial structural organization in the network and seeks to identify a partition maximally different from this model through optimizing the modularity quality function. The I-louvain method modifies the standard modularity quality function to what they label `inertia-based modularity', incorporating a Euclidean distance between nodes based on their attributes, and demonstrating with multiple examples how incorporating connectivity and attributes allows for a partition of nodes to communities that aligns better with ground truth than that obtained using connectivity or attributes in isolation.

Alternatively, there a variety of probabilistic approaches to handling attributed network data \cite{clauset,hric,peel2017ground,cesna}. Similar to our work in the sense that community membership is related to node attributes is CESNA \cite{cesna}. The objective in this approach is to learn a set of propensities or affiliations for each node across all possible communities, such that two nodes with similar propensities towards communities should have more in common in terms of connectivity and attributes. In this model, each node has a vector with multiple binary attributes. The affiliation model is useful and flexible because it does not enforce a hard partitioning of nodes into communities, which is useful in social network applications. In this inference problem, the connectivity and attribute information are used to infer a node's affiliations to communities and then models the probability of an edge between two nodes as a function of the similarity in their community affiliation propensities. 

In contrast to the affiliation model, the stochastic block model \cite{sbmOrig} (at least the more standard variants of it), seeks to determine a hard partition of nodes across communities and models edges between a pair of nodes according to their community assignments. The partition of nodes to communities through a stochastic block model framework is accomplished through maximum likelihood optimization. A variant of the stochastic block model explored by Clauset \emph{et al.}, \cite{clauset} adapts the classic stochastic block model to handle a single attribute with the assumption that attributes (referred to as `metadata') and communities are correlated. Hric \emph{et al}.~\cite{hric} developed an attributed SBM from a multilayer network perspective, with one layer modeling relational information between attributes and the other modeling connectivity, then assigning nodes to communities maximizing the likelihood of the observed data in each layer. Finally, work by Peel \emph{et al.} made important contributions in 1) establishing a statistical test to determine whether attributes actually correlate with community structure and 2) developing an SBM with flexibility in how strongly to couple attributes and community membership in the stochastic block model inference problem \cite{peel2017ground}. 

The model that we seek to develop in this work is distinguished by its ability to fit a stochastic block model to networks where each node has multiple continuous attributes. This model is most appropriate for circumstances where there is domain-specific evidence that members of a community should exhibit similarities in the attributes. We highlight two such examples in section 5, where we apply our model to a protein interaction network and a microbiome subject similarity network. Before discussing these examples, we first define our attributed SBM and an inference technique for fitting the model. We test this approach on a synthetic example. Since community detection methods are often difficult to validate due to the lack of ground truth information on the nodes, we describe the tasks of link prediction and collaborative filtering to quantify how well the attributed SBM represents the data. We then consider these tasks on two biological network examples.

\subsection{Stochastic Block Models}
Because our model is an extension to the widely-used stochastic block model \cite{sbmOrig}, we provide a brief introduction here. This model assumes that edges within a community are connected within and between communities in a characteristic or probabilistic way. To fit this model to network data, the objective is to partition the nodes into communities such that these assignments maximize the likelihood of the model according to the observed edges. In this inference problem for a network with $N$ nodes and $K$ communities, one learns a $K \times K$ probability matrix, ${\boldsymbol \theta}$, describes the probability of connections within and between communities, and an $N$-length vector of node-to-community assignments, ${\bf z}$. For a network with $N$ nodes, $K$ communities, adjacency matrix, ${\bf A}=\{a_{ij}\}$ and a learned vector ${\bf z}$ of node-to-community assignments, the SBM without degree correction (degree-corrected versions also exist \cite{degreeCorrect}) models an edge between nodes $i$ and $j$ with

\begin{equation}
P(a_{ij}=1) \sim \text{Bernoulli}(\boldsymbol \theta_{z_{i}z_{j}})
\end{equation}

The node-to-community assignments (${\bf z})$ are inferred together with the matrix ${\boldsymbol \theta}$ through likelihood optimization. Effective inference techniques for standard stochastic block model parameters are well explored \cite{comparative,tiagomcmc,dudin}, including algorithms for EM, belief propagation, and MCMC accept-reject sampling. 

\section{Model}
\subsection{Objective}
\begin{figure}
\begin{center}
\includegraphics[scale=0.5]{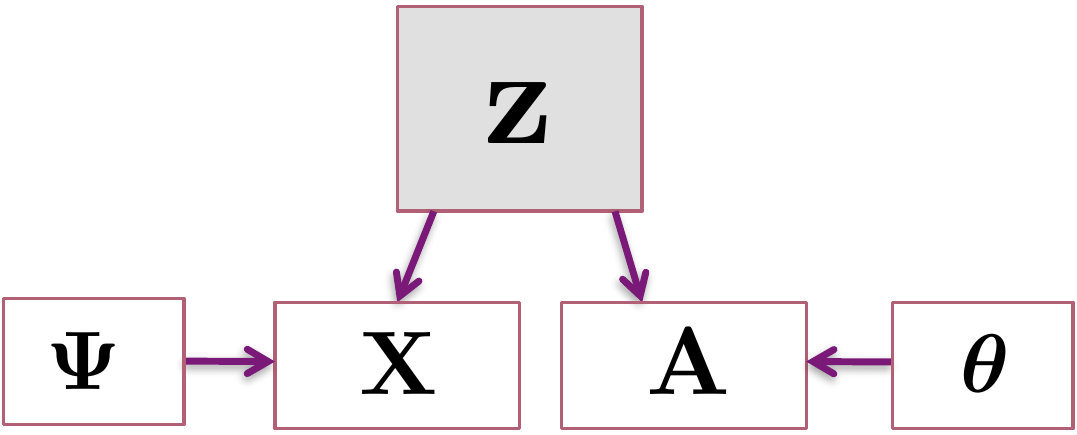}
\caption{{\bf Modeling community membership in terms of attributes and connectivity}. Node-to-community assignments specified by ${\bf Z}$ are determined in terms of adjacency matrix information, ${\bf A}$ and attribute matrix information, ${\bf X}$. ${\bf A}$ and ${\bf X}$ are assumed by be generated from a stochastic block model and a mixture of multivariate Gaussian distributions, parameterized by ${\boldsymbol \theta}$ and ${\boldsymbol \Psi}$, respectively.
\label{fig:graphical_model}} 
\end{center}
\end{figure}

We seek to incorporate both connectivity (${\bf A}$) and attribute information (${\bf X}$) to infer node-to-community assignments, ${\bf Z}$. Note that for a network with $N$ nodes, $K$ communities and $p$ measured attributes, ${\bf A}=\{a_{ij}\}$, ${\bf X}=\{x_{ip}\}$, and ${\bf Z}=
\{z_{ik}\}$ have dimensions $N \times N$, $N \times p$ and $N \times K$, respectively. In particular, we distinguish ${\bf Z}$ to be a binary indicator matrix, where entry $Z_{ic}$ is 1 if and only if node $i$ belongs to community $c$, whereas we also define ${\bf z}$ to be the equivalent $N$-dimensional array labeling node-to-community assignments. We assume connectivity and attributes are conditionally independent, given the community membership label. The graphical model for the relationship between node-to-community labels, connectivity and attribute information is shown in Figure \ref{fig:graphical_model}.

To infer the ${\bf Z}$ that best explains the data, we adopt a likelihood maximization approach. That is, we seek to find the partition of nodes to communities that best describes the observed connectivity and attribute information.  Given the conditional independence assumption of ${\bf X}$ and ${\bf A}$, we can express the log likelihood of the data, ${\mathcal L}$ as the sum of connectivity and attribute log likelihoods, ${\mathcal L}_{A}$ and ${\mathcal L}_{X}$, respectively,
as
\begin{equation}
\mathcal{L}=\mathcal{L}_{A}+\mathcal{L}_{X}\enspace.
\label{eqn:likelihood_decomposition}
\end{equation}

This likelihood reflects the joint distribution of the adjacency matrix, ${\bf A}$, the attribute matrix, ${\bf X}$, and the matrix of node-to-community indicators, ${\bf Z}$; formally, we have 

\begin{equation}
\mathcal{L}=p({\bf A},{\bf X},{\bf Z})\enspace. 
\end{equation}

Given that ${\bf Z}$ is a latent variable that we are trying to infer, we can approach the problem using the expectation maximization (EM) algorithm \cite{EM}. By doing this, we will alternate between estimating the posterior probability that a node $i$ has community label $c$, or

\begin{equation}
\label{post}
p(z_{ic}=1\mid {\bf X,A})
\end{equation} 
and estimates for ${\boldsymbol \theta, \boldsymbol \Psi}$, i.e., the model parameters specifying the adjacency and attribute matrices, respectively. 

\subsection{Attribute Likelihood}

For a network with $K$ communities, we assume that each particular community $i$ has an associated $p$-dimensional mean ${\boldsymbol \mu}_{i}$ and $p \times p$ covariance matrix, ${\boldsymbol \Sigma}_{i}$. Note that these parameters uniquely identify a $p$-dimensional multivariate Gaussian distribution. To specify this model for all $K$ communities, we define the parameter ${\boldsymbol \Psi}=\{{\boldsymbol \mu}_{1},{\boldsymbol \mu}_{2},\dots {\boldsymbol \mu}_{k},{\boldsymbol \Sigma}_{1},{\boldsymbol \Sigma}_{2},\dots {\boldsymbol \Sigma}_{K}\}$. 

The log likelihood for the mixture of Gaussians on the attributes is written as,

\begin{equation}
P({\bf X} \mid {\boldsymbol \Psi})=\sum_{i=1}^{N}\log\{\sum_{c=1}^{K}\pi_{c}\mathcal{N}({\bf x}_{i}\mid {\boldsymbol \mu}_{c},{\boldsymbol \Sigma}_{c})   \}
\end{equation}

%
Here, $\mathcal{N}({\bf x}_{i} \mid {\boldsymbol \mu}_{c},{\boldsymbol \Sigma}_{c})$ is the probability density function for the multivariate Gaussian and $\pi_{c}$ is the probability that a node is assigned to community $c$.

\subsection{Adjacency Matrix Likelihood}
For the adjacency matrix, ${\bf A}$ and the $K \times K$ matrix of stochastic block model parameters, ${\boldsymbol \theta}$, the complete data log likelihood can be expressed as

\begin{equation}
\begin{split}
\log(P({\bf A} \mid {\bf z}))&=\frac{1}{2}\sum_{i\ne j}\sum_{k,l}z_{ik}z_{jl}[a_{ij}\log (\theta_{kl})\\
&+(1-a_{ij})\log(1-\theta_{kl})]\enspace. 
\end{split}
\end{equation}



\subsection{Inference}
To use EM to maximize the likelihood of the data, we break the process into the E-step and M-Step, and perform this step sequence iteratively until the estimates converge.

\textbf{E-Step.} During the E-step, we use the current value of learned model parameters, ${\boldsymbol \theta}$ and ${\boldsymbol \Psi}$ to compute the posterior given in Eq.~\eqref{post} at each step. The posterior at each step,  $\gamma(z_{ic})$, of node $i$ belonging to community $c$, is given by

\begin{equation}
\begin{split}
\gamma({z_{ic}})& =p(z_{ic}=1\mid {\bf x}_{i}, {\bf a}_{i}) \\
&=\frac{p({\bf x}_{i} \mid z_{ic}=1)p({\bf a}_{i}\mid z_{ic}=1)\pi_{c}}{\sum_{c=1}^{K}p({\bf x}_{i} \mid z_{ic}=1)p({\bf a}_{i}\mid z_{ic}=1)\pi_{c}}\enspace.
\end{split}
\end{equation}

Here, ${\bf x}_{i}$ and ${\bf a}_{i}$ denote the attribute and connectivity patterns for node $i$, respectively. 

\vskip1ex
\textbf{M-Step.} In the M-step, we can compute updates for ${\boldsymbol \theta}$ and ${\boldsymbol \Psi}$ using this expectation.

Since, the attributes follow a Gaussian mixture model, it can be shown that the update for the mean vector describing community $c$, ${\boldsymbol \mu}_{c}$, can be computed as

\begin{equation}
{\boldsymbol \mu}_{c}=\frac{\sum_{i=1}^{N}\gamma(z_{ic}){\bf x}_{i}}{\sum_{i=1}^{N}\gamma(z_{ic})}\enspace.
\end{equation}

Similarly, the update for the covariance matrix describing a community, ${\boldsymbol \Sigma}_{c}$,  is computed as 

\begin{equation}
{\boldsymbol \Sigma}_{c}=\frac{\sum_{i=1}^{N}\gamma(z_{ic})({\bf x}_{i}-{\boldsymbol \mu}_{c})({\bf x}_{i}-{\boldsymbol \mu}_{c})^{T}}{\sum_{i=1}^{N}\gamma(z_{ic})}\enspace.
\end{equation}

To update the parameters of ${\boldsymbol \theta}$, we follow the method in \cite{dudin} and update the probability of an edge existing between community $q$ and $l$, given by ${\boldsymbol \theta}_{ql}$ as,

\begin{equation}
{\boldsymbol \theta}_{ql}=\frac{\sum_{i\ne j} \gamma(z_{iq})\gamma(z_{jl})x_{ij}}{\sum_{i \ne j}\gamma(z_{iq})\gamma(z_{jl})}
\end{equation}

We continue the process of iterating between the E-step and M-step until the change in the data log-likelihood, ${\mathcal L}$, is below a predefined tolerance threshold.

\subsection{Initialization}
Likelihood optimization approaches are often sensitive to initialization because it is easy to get stuck in a local optimum. As an initialization strategy for the nodes, we simply cluster the nodes in the network using the Louvain algorithm \cite{louvain}. We chose this approach because this algorithm is efficient and stable.  

\section{Synthetic Data Results}
We first test the performance of our model and inference procedure on a synthetic example. We generated networks with a stochastic block model with $N=200$ nodes and $K=4$ communities, parameterized as follows: 
\begin{equation}
p(A_{ij}=1) \sim 
\begin{cases}
~\text{Bernoulli}(.10), & \mbox{if } z_{i} \ne z_{j}\\
~\text{Bernoulli}(.25), & \mbox{if } z_{i} = z_{j}\\
\end{cases}
\end{equation}
Note that ${\bf z}$ is a 200-dimensional vector, where $z_{i}$ identifies the community label for node $i$. 

\begin{figure}[h!]
\begin{center}
\includegraphics[width=0.5\textwidth]{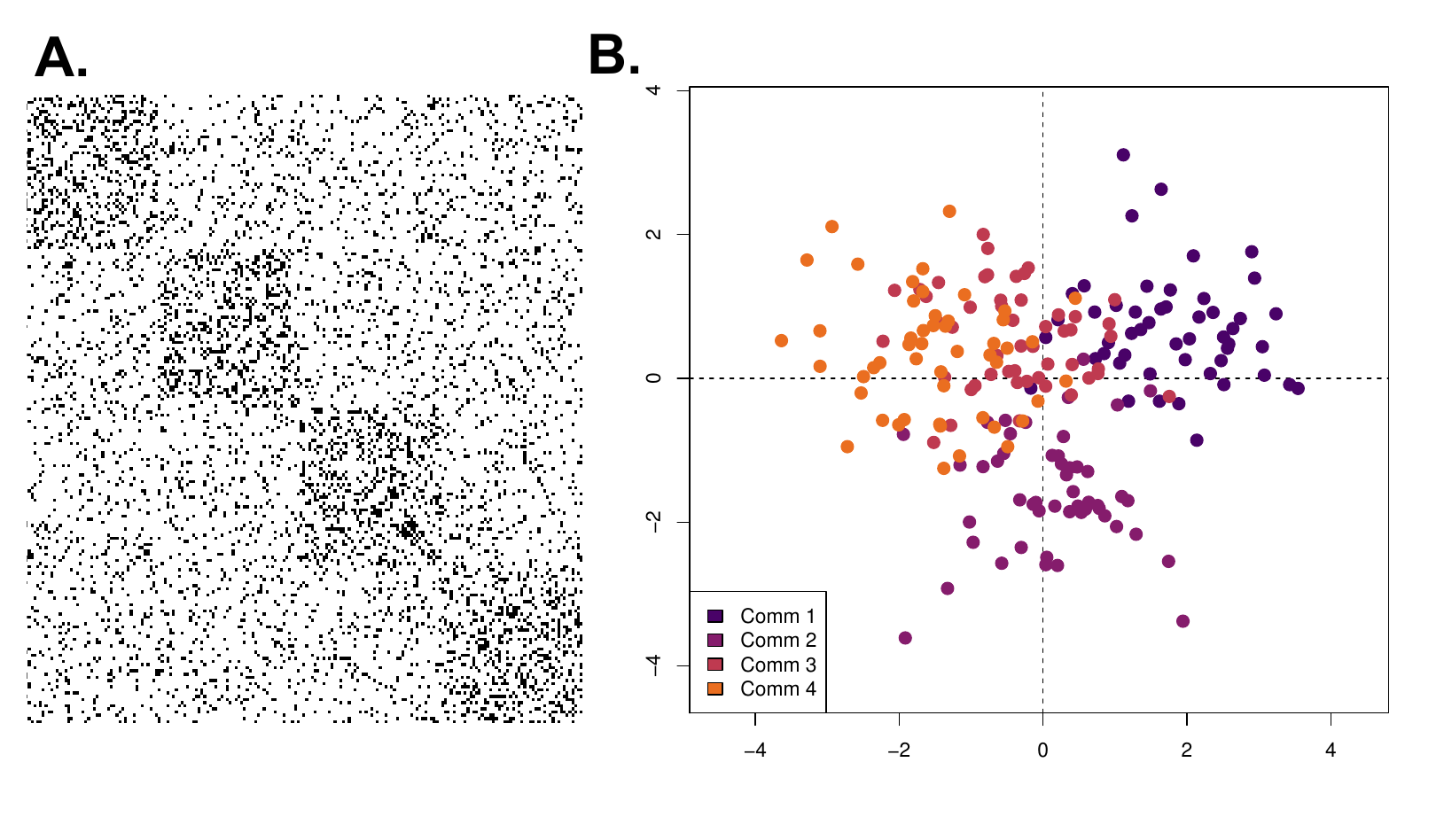}
\caption{{\bf Synthetic Example.} We generated a synthetic network with $N=200$ nodes, $K=4$ communities and an 8-dimensional multivariate Gaussian for each community. {\bf A.} A visualization of the adjacency matrix for this network where a black dot indicates an edge. We observe that there is an assortative block structure (blocks on the diagonal), but there are also many edges between communities making the true community structure using only connectivity harder to detect. {\bf B.} We performed PCA on the $N \times p$ attribute array and plotted each of the $N$ nodes in two dimensions. Points are colored by their true community assignments, ${\bf z}$. Clustering the nodes according to only connectivity, only attributes, and with the attributed SBM, we quantified the partition accuracy with normalized mutual information, yielding $\text{NMI}({\bf z},\{{\bf z}^{\text{connectivity}}, {\bf z}^{\text{attributes}},{\bf z}^{\text{attribute sbm}}\})=\{0.65,0.68,0.83\}$.}
\label{Fig2}
\end{center}
\end{figure}

\vskip1ex
Figure \ref{Fig2}A shows the adjacency matrix for an example network generated according to this parametrization. The black marks in the image indicate an edge. While this network has assortative structure with members of a community having more edges on average with each other than with other communities, there are still many noisy edges going between communities, making the correct community structure more difficult to discern. 

To model attributes, for a community $c$, we randomly generated an 8-dimensional vector, $\boldsymbol{\mu}_c$, where each entry is from a Gaussian with 0-mean and unit variance. Associated with each $c \in \{1,2,3,4\}$ is an $8 \times 8$ diagonal covariance, ${\boldsymbol \Sigma}_{c}=\text{diag}(1.25)$. Moreover, using the ${\boldsymbol \mu}_{c}$ and ${\boldsymbol \Sigma}_{c}$, a sample attribute vector can be generated. That is, the attribute vector ${\bf x}_{i}$ for node $i$ is generated as
\begin{equation}
{\bf x}_{i} \sim \mathcal{N}({\boldsymbol \mu}_{z_{i}},{\boldsymbol \Sigma}_{z_{i}})
\end{equation}
where $\mathcal{N}(\cdot,\cdot)$ denotes a multivariate Gaussian.

\vskip1ex
Figure \ref{Fig2}B shows a PCA plot of the attribute vectors associated with each node in an example synthetic experiment, with each point representing a node. Since the true dimension of these feature vectors is 8, this plot provides a projection onto the first 2 principal components, allowing a visualization of the relatedness between node attributes. One can observe that members of community 2 are overall nicely separated from other communities in the projected attribute space but members of communities 3 and 4 are especially hard to discern here. \\

To assess how well the attribute SBM approach performed in successfully assigning nodes to communities, we compared the results obtained from our model to clustering results obtained clustering based only on connectivity and to clustering based only on the attribute information. We quantify the correctness of the obtained partitions with normalized mutual information (NMI) \cite{commdeccompare}. Letting ${\bf z}$ denote the true node-to-community assignments, then ${\bf z}^{\text{connectivity}}$, ${\bf z}^{\text{attributes}}$, and ${\bf z}^{\text{attribute sbm}}$ denote the partition of the nodes according to the network connectivity only, attributes only, and with the attributed SBM. To cluster the network only according to connectivity, we fit a stochastic block model with 4 blocks. To cluster nodes with only attributes, we performed $k$-means clustering on only the attributes. Computing the NMI between ${\bf z}$ and each of these 3 cases, we obtain 0.65, 0.68, and 0.83, respectively. 
These results show that by combining both sources of information, there is an improvement in the ability to correctly identify communities. To further probe this idea, we sought to empirically look closer at the so-called 'detectability limit'. Generally, detectability refers to the difficulty of correctly identifying clusters in data; in particular, sharp phase transitions are observed in fitting stochastic block models, with accurate capture of the correct communities only if the within-community probability, $p_{in}$, is sufficiently larger than the between-community probability, $p_{out}$ \cite{decelle,taylor}. 

\begin{figure}[h!]
\begin{center}
\includegraphics[width=0.4\textwidth]{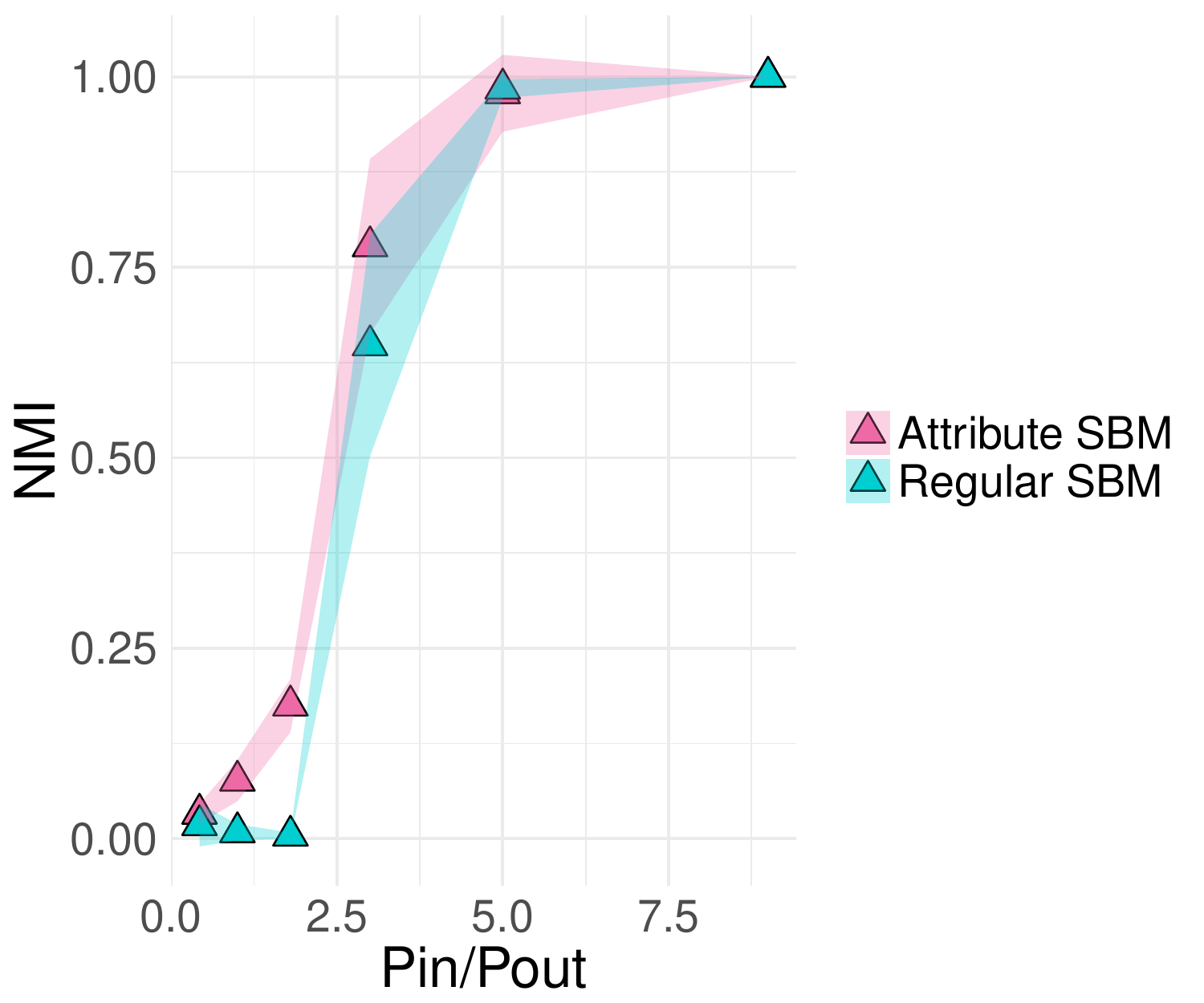}
\caption{{\bf Detectability Analysis in Synthetic Example.} To understand how attribute information can be combined with connectivity to assign nodes to communities accurately, we generated synthetic networks for within-probabilities of $p_{in}$ between 0.05 and 0.3 with corresponding $p_{out}$ or between-community probabilities such that the mean degree of the network was 20. For each of these synthetic networks, we used the attributes from the analysis in figure 2 to fit the attributed SBM. Here, we plot the correctness of the node-to-community assignment with normalized mutual information using the partition obtained from regular SBM (blue) and the partition under the attributed SBM model fit (pink). For each combination of $p_{in}$ and $p_{out}$, we generated 10 networks and hence the bands around the points denote standard deviation. Incorporating attributes with the attributes stochastic block model improves results, particularly near and below the detectability limit, and appears to smooth out the sharp phase transition. }
\label{Fig3}
\end{center}
\end{figure}

Based on the results of the synthetic experiments in Figure \ref{Fig2} where the attributes combined with connectivity lead to a more accurate partitioning of the nodes, we hypothesized that augmenting the network connectivity with attributes may move this detectability limit. In Figure \ref{Fig3}, we explored how generating networks from a stochastic block model with varying ratios between $p_{in}$ and $p_{out}$ combined with the attributes used in Figure \ref{Fig2} would affect the accuracy of the node-to-community partition. To do this, we considered values of $p_{in}$ between 0.05 and 0.3 in increments of 0.05. For each of these $p_{in}$ values, we found the corresponding value of $p_{out}$ such that the mean degree was 20. Fixing the mean degree allows for direct comparison of how the within-to-between community probabilities influence the detection of correct communities. For each of these $p_{in}$ and $p_{out}$ combinations, we generated 10 different networks using a stochastic block model. In figure 3 we plot the NMI between the true partition, ${\bf z}$ and the partitions using only the connectivity with the regular SBM ${\bf z}^{\text{connectivity}}$ and the attributed SBM ${\bf z}^{\text{attribute sbm}}$. These results are plotted in blue and pink, respectively. The shaded region around the points indicates standard deviation. 

We see that while both inference approaches undergo a strong increase in accuracy at a similar ratio of $p_{in}/p_{out}=3$, we notice that the curve for the attribute SBM results are slightly shifted to the left due to the use of the extra attribute information positively impacting the ability to correctly identify communities. Moreover, we note that the attribute SBM results appear to smooth out the sharp phase transition that is visible in the results from the SBM without attributes. Future work could focus on better understanding the impact on such detectability questions in terms of the parameters for the underlying multivariate Gaussian distributions parametrizing each community. 


\section{Using the fitted attributed SBM for link prediction and collaborative filtering}
One of the benefits of a generative network model is that it can be applied to prediction tasks. Most notably, in the absence of one source of information about a node (connectivity or attributes), the model can be used to predict the complementary information source (attributes or connectivity, respectively). We demonstrate here that fitting an attributed SBM may provide a means to successfully perform two fundamental network prediction tasks: link prediction and collaborative filtering. 

In the link prediction problem, when given two node stubs, the objective is to determine whether a link exists between them. Since we are modeling connectivity with a stochastic block model, we can predict links using the learned parameters.  In particular, we highlight how this task can be performed using just the attribute information of the node stubs of interest. In the experiments to follow, we compare to 3 commonly-used link prediction methods. In all of these methods, a score is computed for all edge-candidate dyads and ultimately the top $x$ set of prospective edges with highest weights are kept (where $x$ is some user-defined parameter). Let $m$ and $n$ be a pair of nodes and $\Gamma(m)$ denote the set of neighbors for a node $m$. Then, under the following 3 common  link prediction methods \cite{linkPredReview}, we can calculate the score of the potential link as $\text{Score}(m,n)$.

{\bf Jaccard}: $\text{Score}(m,n)= \frac{\Gamma(m) \cap \Gamma(n)}{\Gamma(m)\cup \Gamma(n)}$

{\bf Adamic Adar}: $\text{Score}(m,n)=\sum_{c \in \Gamma(m) \cap \Gamma(n)}\frac{1}{\log |\Gamma(c)|}$

{\bf Preferential Attachment}: $\text{Score}(m,n)=|\Gamma(m)|\times |\Gamma(n)|$\\

Conversely, the collaborative filtering problem seeks to predict a node's attributes based on its similarity to its neighbors. For some node of interest, we can use our fitted attributed SBM model to predict a node's attributes, given only the information about its connectivity. Formally, for node $i$, we seek to predict ${\bf x}_{i}$. In the following experiments, we compare our results to two common collaborative filtering approaches \cite{collabFilterReview}. Let $\mathcal{N}^{k}(m)$ be the set of $k$-nearest neighbors in the network for node $m$. Let $\hat{{\bf x}_{i}}$ be the predicted attribute vector for node $i$ and $s_{ij}$ be a similarity measure between nodes $i$ and $j$. 

{\bf Neighborhood Avg}: $\hat{{\bf x}_{i}}= \frac{1}{|\mathcal{N}^{k}(i)|}\sum_{j \in \mathcal{N}^{k}(i)} {\bf x}_{j}$

{\bf Weighted Neighborhood Avg}: \[\hat{{\bf x}_{i}}=\frac{1}{\sum_{j in \mathcal{N}^{k}(i)}s_{ij}}\sum_{j \in \mathcal{N}^{k}(i)}s_{ij}{\bf x}_{j}\]

We show results for these two tasks in two different biological network examples in section 5. In particular, the experiments were designed in the following ways.

\subsection{Link Prediction Experiments}
For the link prediction tasks shown in Figures \ref{Fig5} and \ref{Fig9}, we performed a link prediction task by sampling pairs of nodes and utilizing the complementary source of attribute information. We sampled 10 different sets of 50 pairs of nodes. In each sample, 25 of the node pairs were those having an edge in the original network and 25 were pairs with no edge. For each of the 50 edges in each sample, we sought to predict whether an edge existed between the corresponding node pair in a leave one out manner. To do this, for each edge we fit the attributed SBM to the network with the pair of nodes (stubs) associated with the edge removed. We then use the nearest neighbor in attribute space of each stub as the the input to each of the 3 baseline community detection methods (Jaccard, Adamic Adar, and Preferential Attatchment). To use our attributed SBM in this link prediction task, we also consider the most commonly observed community among the 3 nearest neighbors for the stubs of the edge of interest. Again, using the nearest neighbors, which we denote by $n$ and $m$ of the stubs, then we define the link prediction score for the edge as $\theta_{z_{n},z_{m}}$, or the probability that an edge exists between nodes $n$ and $m$ according to the fitted model. After generating 10 samples of 50 edge pairs, this results in 500 total edge scores. Since we know the ground truth of whether or not these edges actually exist from the original network, we can construct an ROC curve for each method. From these curves we can plot area under the curve (AUC) to quantify the quality of the link prediction result. Using the attributed SBM is a way to incorporate community information into the link prediction problem which has previously shown to be effective \cite{linkComm}.

\subsection{Collaborative Filtering Experiments}
In collaborative filtering experiments, the objective is to predict the vector of attributes for each node. In our experiments, we used leave-one-out validation to predict the attribute vector for each node. That is, for each node in the network, we created a single node test set. The training set, was then the rest of the network with the node to predict removed. For this single test set node, we identified neighbors it connects to in only connectivity space within the training set. For standard collaborative filtering approaches (Neighborhood average and weighted neighborhood average), the predicted attribute for the test set node is then the specified averaging of the neighbors. To use our model for this task, we first fit the attributed SBM model to the training set. Similar to the standard link prediction approaches, we identify the nearest neighbors for our test node in connectivity space within the training set. We then predict the community membership of our test node to be the most-frequently observed community among its neighbors. Using this community assignment, $c$, we then predict the attribute vector, ${\bf x}_{i}$ for a node $i$. These results are Figures \ref{Fig6} and \ref{collabprotein}. For a node $i$ and its associated vector of attributes, ${\bf x}_{i}$ we quantify the accuracy of the predicted attribute vector, $\hat{{\bf x}_{i}}$ with a a relative error measure, $\mathcal{E}$, such that 
\begin{equation}
\mathcal{E}=\frac{||\hat{\bf{x}}_{i}-{\bf x}_{i}||_{2}}{||{\bf x}_{i}||_{2}}.
\end{equation}

Similar to the success of integrating community information for link prediction, collaborative filtering tasks have previously shown success from the integration of network community structure \cite{collabComm}. 

\section{Applications in Biological Networks}
We evaluate the potential to combine similarity or relational information between a set of entities for application in biological data. For example, one might consider networks of proteins, genes, or bacterial species with extra experimental data. Our application of this model to biological problems provides a framework to predict attribute or connectivity information about a new observation.  Note that we do not intend to suggest any new biological insights here, but rather that we can combine two sources of information for prediction tasks and alternative definitions of what constitutes a community in the data. Applying the attributed stochastic block model to integrate connectivity and attribute data provides a way to find a partition that takes into account two different sources of information, or a method to predict one source of information (connectivity, attributes) in the absence of the other (attributes, connectivity). 

\subsection{Microbiome Subject Similarity Results}

{\bf Motivation}

In the analysis of biological data, it is often useful to cluster subjects based on a set of their measured biological features and to then determine what makes each of the subgroups different. One type of biological data gaining much attention in recent years is metagenomic sequencing data, used to profile the composition of a microbiome. We refer to this as the 'metagenomic profile' and each feature is a count for each bacterial species, also known as operational taxonomic unit (OTU). Lahti \emph{et al}.\ conducted a study among subjects across a variety of ethnicities, body mass (BMI) classifications, and age groups to understand differences in the intestinal microbiota \cite{microbiomedata}. Using metagenomic sequencing, the counts for 130 OTUs were provided for each subject. We created an experiment to test our model by seeing if we could overlay a similarity network between subjects with the individual OTU count vectors for each subject. 

{\bf Pre-Processing}
The data were downloaded from \url{http://datadryad.org/resource/doi:10.5061/dryad.pk75d}. We extracted a subset of the subjects from Eastern Europe, Southern Europe, Scandinavia, and the United States. Using only these subjects, a between-subject similarity network was constructed between the 121 individuals who had a BMI measurement. This resulted in a network of 121 nodes, where each edge is the Pearson correlation between their microbial compositions. We then removed all edges in the network with weight (correlation) $<0.7$. Note that our attributed SBM does not allow for edge weights, so we simply ignored the edge weights as input to the model.

{\bf Constructing Node Attributes}
Since each node had a 130-dimensional vector of attributes (counts), we used this information to create a lower-dimensional attribute vector for each node by performing PCA and then representing each node with the first 5 principal components. Each dimension of this new attribute vector was then centered and scaled, and we observed an approximately Gaussian distribution. 

\begin{figure}[h!]
\begin{center}
\includegraphics[width=0.5\textwidth]{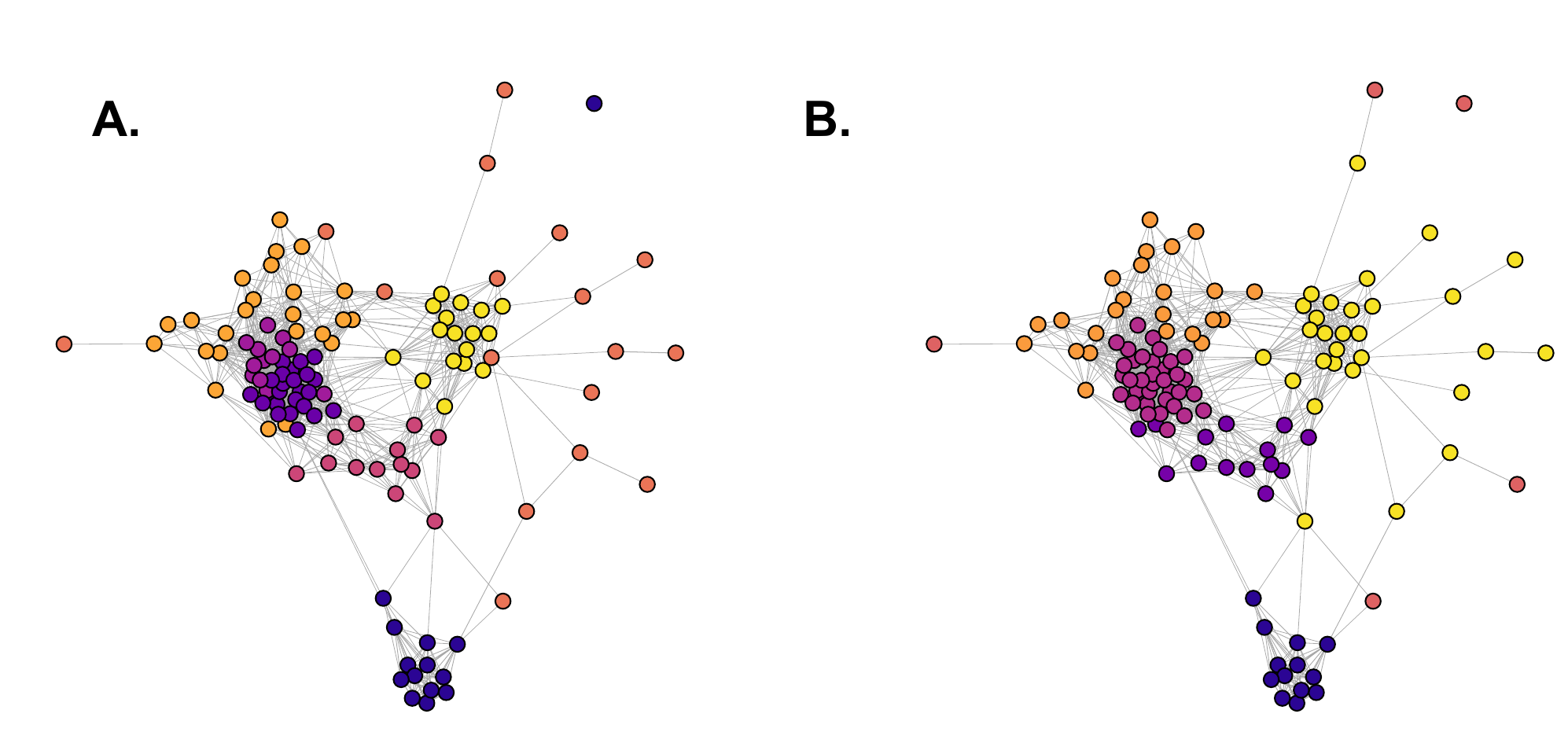}
\caption{{\bf Microbiome subject similarity network:} A visualization of the 121 node microbiome subject similarity network with nodes colored by the partition using the classic ({\bf A.}) and attributed ({\bf B.}) stochastic block model. {\bf A.} Fitting the classic stochastic block model to the network, 7 communities were identified. {\bf B.} Fitting the attributed stochastic block model to the network with the attributes being the first 5 principle components of each subject's OTU count vector (metagenomic profile), 6 communities were identified. Incorporating attributes in inferring this partition removed some of the noise in the partition on the network, specifically in the mixed purple community in the left of {\bf A.}}
\label{Fig4}
\end{center}
\end{figure}

We first visualized the differences in partitions obtained according to the classic and attributed stochastic block models in Figure \ref{Fig4}A-B, respectively. In both networks, nodes are colored by their community assignment. Using the classic stochastic block model and the model selection criteron described in \cite{dudin}, 7 blocks were identified. With the attributed stochastic block model, 6 blocks were identified. While we do not have ground truth labels on the nodes, it is visually apparent that adding the attributes to the inference problem helps to `clean up' the partition. For example, in Figure \ref{Fig4}A there is mixing between the dark and lighter purple communities in the upper left of the network. In Figure \ref{Fig4}B, this mixing was reduced by assigning all of the nodes in the general region to the lighter purple community. 

\begin{figure}
\begin{center}
\includegraphics[width=0.4\textwidth]{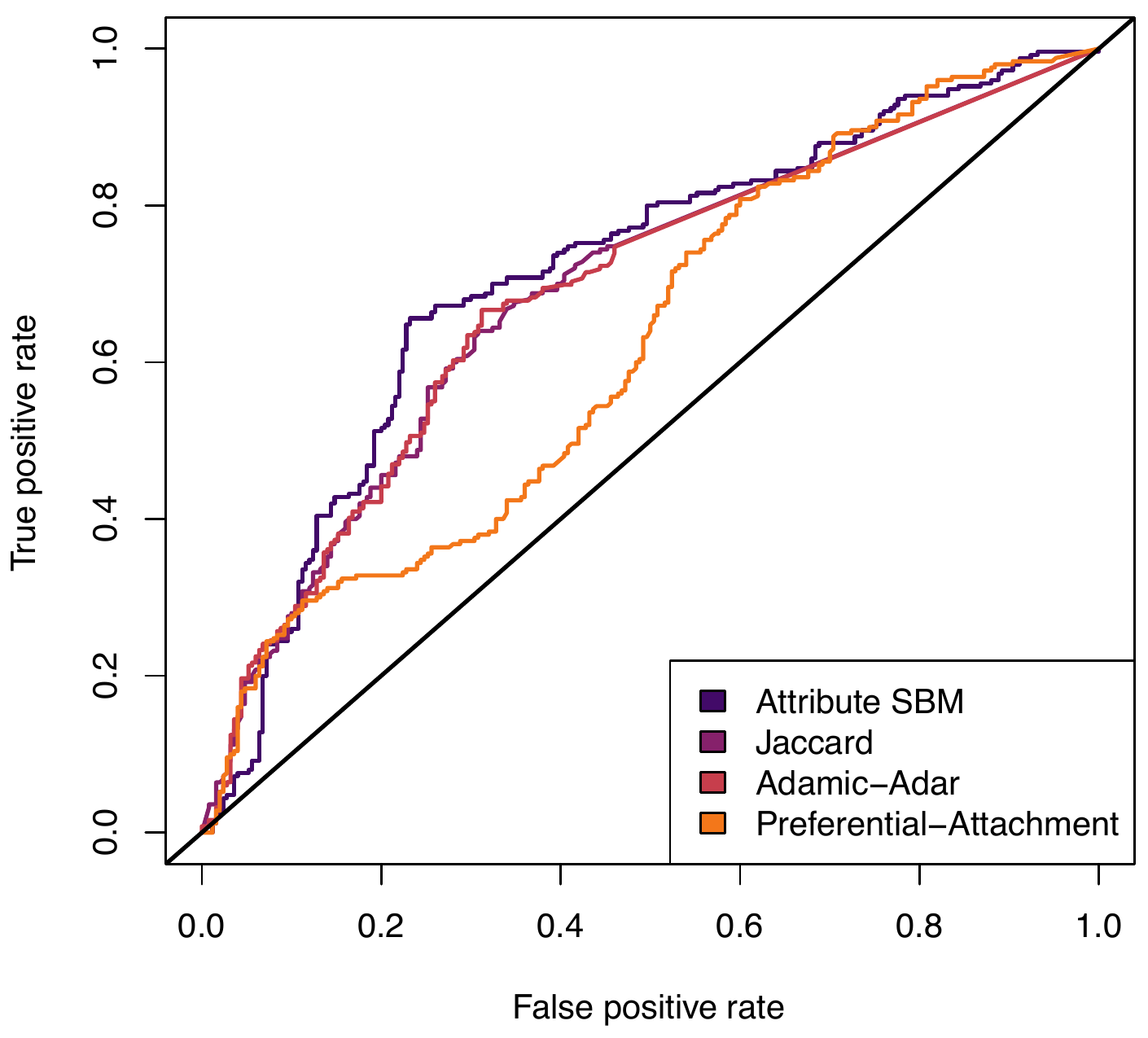}
\caption{{\bf Link Prediction on the microbiome subject similarity network:} The results for link prediction on the microbiome subject similarity network for the attributed SBM, Jaccard, Adamic-Adar and preferential attachment methods. The corresponding AUC values for these methods, respectively are, 0.71, 0.69, 0.69, and 0.62.}
\label{Fig5}
\end{center}
\end{figure}

{\bf Microbiome Link Prediction}
We performed link prediction on the microbiome subject similarity network as described in section 4.1. The associated ROC curves are plotted in Figure \ref{Fig5}. All four methods have satisfactory performance with the attributed stochastic block model giving the best results. The AUC values for the attributed SBM, Jaccard, Adamic-Adar, and preferential attachment are 0.71, 0.69, 0.69, and 0.62, respectively. 
 
\begin{figure}[h!]
\begin{center}
\includegraphics[width=0.3\textwidth]{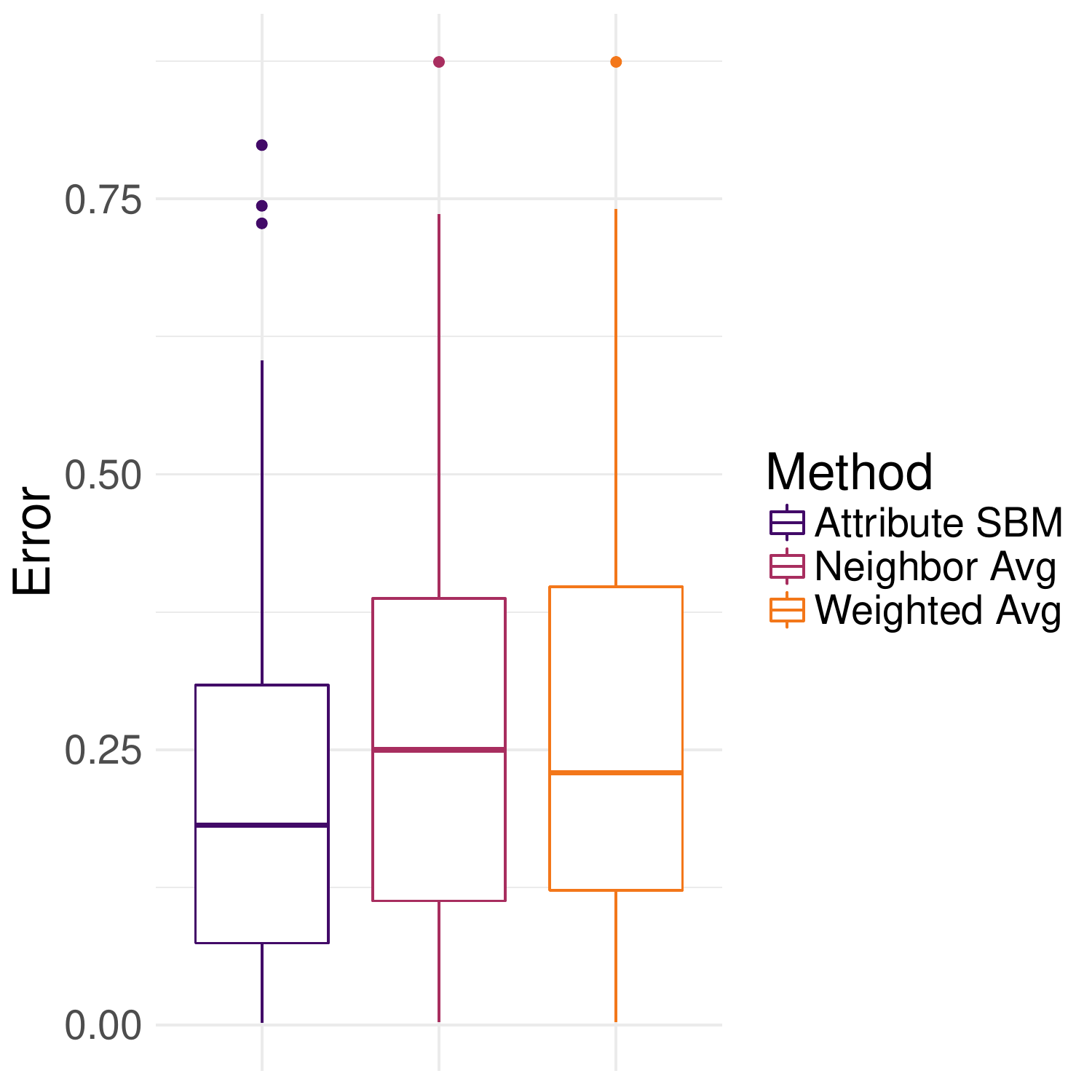}
\caption{{\bf Collaborative Filtering Accuracy in Microbiome Subject Similarity Network}: For each of the 121 nodes, we fit a model to the remaining 120 node network and given the node's closest  neighbors (based on network connectivity) sought to predict its 5-dimensional attribute vector. The reported error is the relative error $\mathcal{E}$ between the difference between the true attribute vector (${\bf x}_{i}$) and its predicted attribute vector (${\hat{\bf x}}_{i}$). The mean error in ${\bf x}_{i}$ is 0.21, as opposed to the neighbor average and weighted neighbor averages, having errors of 0.26 and 0.27, respectively. }
\label{Fig6}
\end{center}
\end{figure}

{\bf Microbiome Collaborative Filtering}
We performed the collaborative filtering experiments on the microbiome subject similarity network in the manner described in section 4.2 to predict the 5-dimensional attribute vector for each node. The box plots in Figure \ref{Fig6} indicate the distribution of relative errors over the 121 nodes for the attribute SBM (blue), neighbor average (pink) and weighted neighbor average (orange). While the attributed SBM plotted has a similar distribution of relative errors with the standard collaborative filtering methods, the mean is slightly lower, at 0.21, compared to 0.26 and 0.27 in the neighbor average and weighted neighbor average, respectively.

\subsection{Protein Interaction Network Results}
We also apply our attributed SBM approach to the protein interaction network presented in \cite{bonacci}. This network represents interactions between proteins, predicted from the literature. Associated with each each node (protein), is a classification of one of 6 experimental modifications observed from the exposure of cancer cells to a chemotherapeutic drug. While communities in this network should reflect functional relatedness among proteins (e.g.\ similar biological functions, in general), we also expect that members of a community should share similarities in the observed modification type. Also associated with each of the 6 modification types is whether that particular type of modification became either more or less prominent after treatment with the drug. Since we have two types of labels associated with these nodes, we also sought to explore how these two labeling schemes (6 class vs. 2 class) aligned with the communities returned by the algorithm. 

\begin{figure}[h!]
\begin{center}
\includegraphics[width=0.5\textwidth]{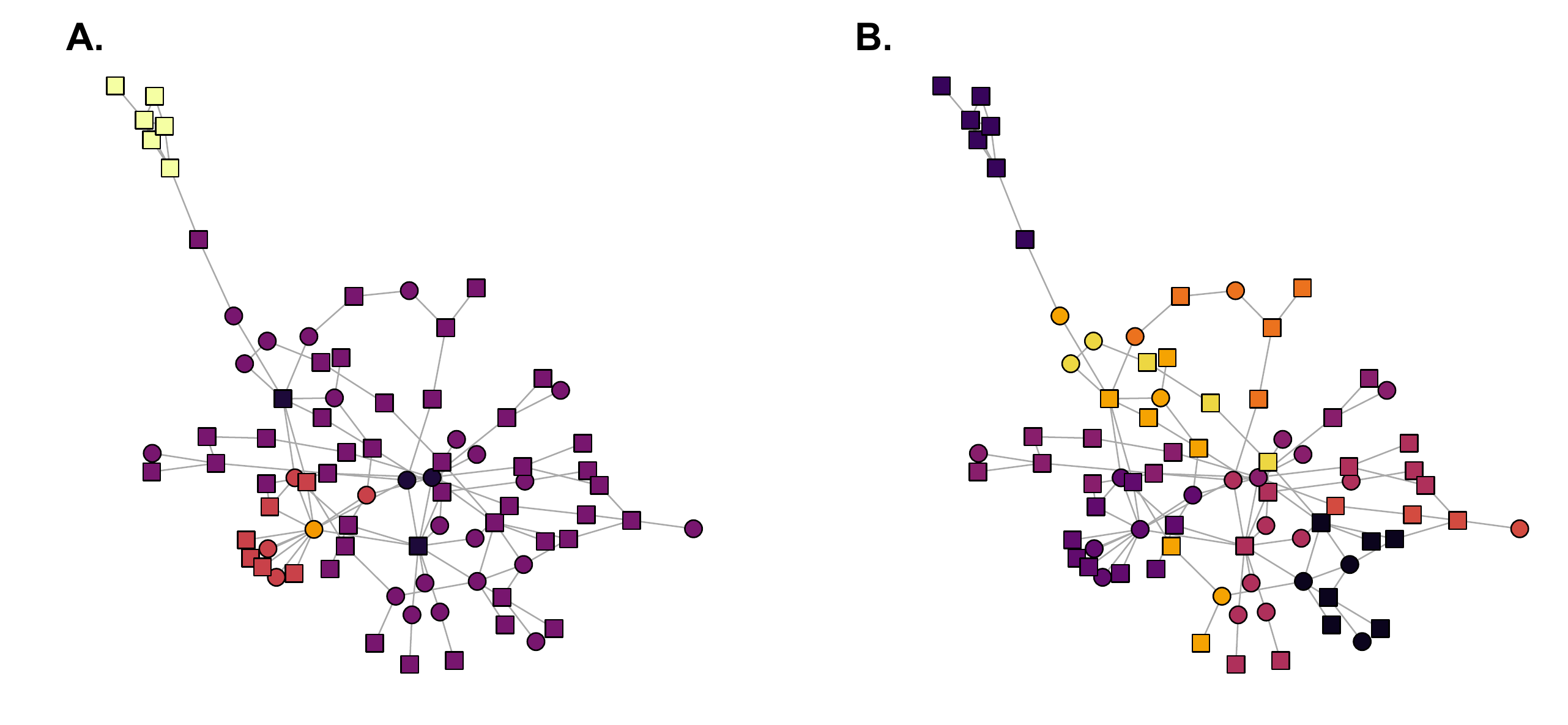}
\caption{{\bf Protein interaction network.} We visualize the 82 node protein interaction network under the classic stochastic block model {\bf A.} and the attributed stochastic block model {\bf B.} In both networks, nodes are colored by their community assignment and the node shape indicates whether the modification status increased (square) or decreased. {\bf A.} Nodes colored according to the community partition under the stochastic block model. Nodes are assigned to one of five communities. {\bf B.} Nodes are colored to the community partition under one of nine communities.}
\label{Fig7}
\end{center}
\end{figure}

{\bf Data Pre-Processing: } We downloaded the unweighted protein interaction network data and the modification information from the supplement of \cite{bonacci}. We removed 13 nodes that were not connected to the largest component of the network and considered only the 82 node largest connected component.

{\bf Constructing Node Attributes:}
Each node is classified with 1 of 6 possible modification types. For each node, we created an attribute vector that captured the modification types of its neighbors. To do this, we considered the 4th order neighborhood of each node. That is, for each node, we collected its neighbors who were four hops or less away in the network. Then to define the value for attribute $c$ of node $i$, or $x_{ic}$, we counted the number of 4th order neighbors of node $i$ with label $c$. After defining these attributes across all nodes, for each of the 6 classes, we centered and scaled each attribute across all of the nodes to have mean 0 and unit variance. 

\begin{figure}[h!]
\begin{center}
\includegraphics[width=0.5\textwidth]{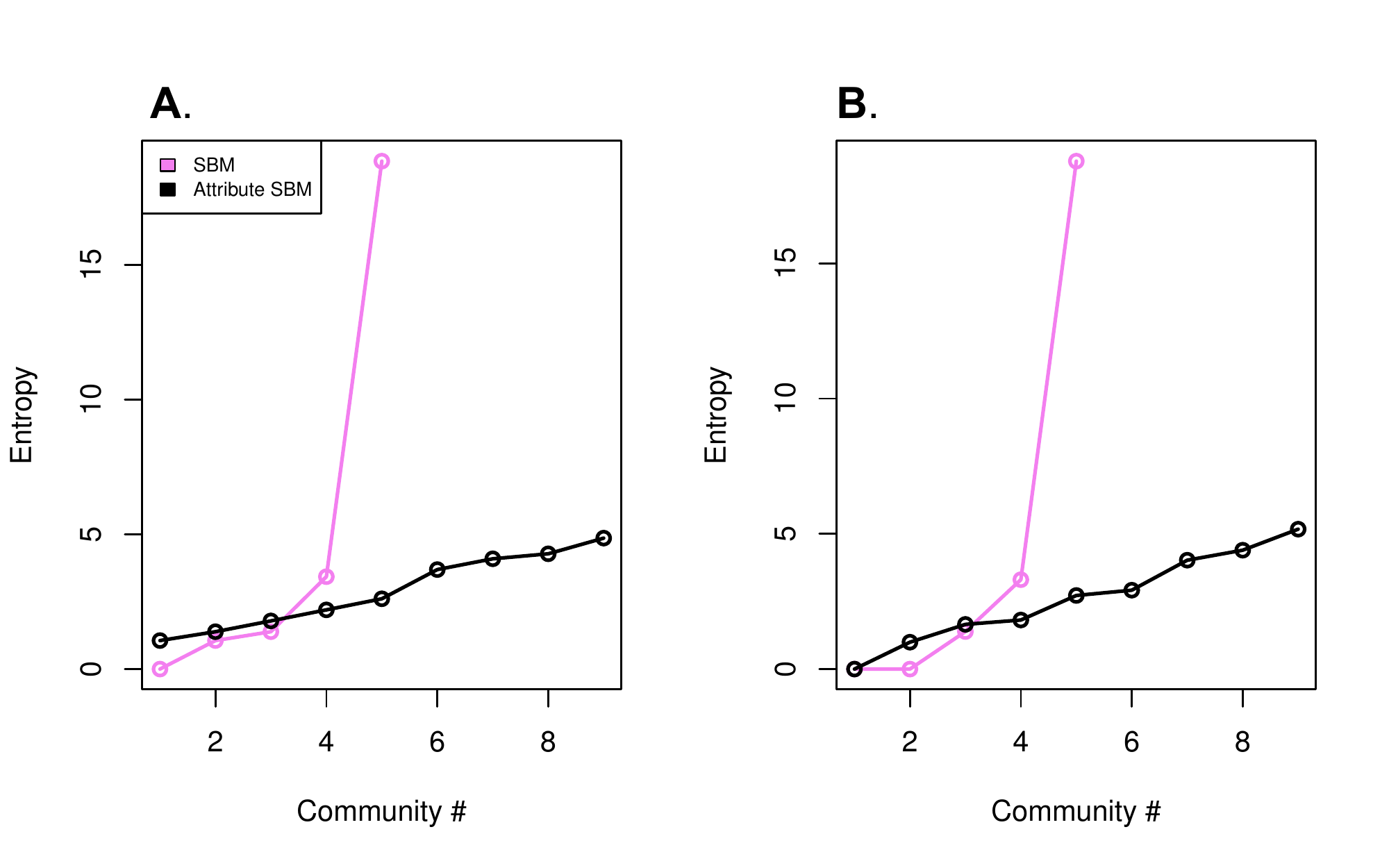}
\caption{{\bf Community entropies in the protein interaction network.} We studied the entropy of the 2 class and 6 class classifications of the nodes in {\bf A.} and {\bf B.}, respectively under the classic SBM (black) and attributed SBM (purple) partitions. For ${\bf A.-B.}$ the horizontal axis denotes the community index for the particular partition. Nodes belonged to 1 of 5 communities under the classic SBM and belong to 1 of 9 communities with the attributed SBM. Incorporating attributes under both classifications succeeds in breaking up a high entropy community (5) from the classic SBM partition to lower entropy communities in the attributed SBM partition. }
\label{entropyFig} 
\end{center}
\end{figure}

Figure \ref{Fig7}A-B show the results of fitting a classic SBM and attributed SBM, respectively. Nodes are colored by their community assignment. The 6 possible modifications arise from 3 biological processes that can either increase or decrease after exposure to the drug. The node shape reflects whether the experimental modification for a node increased (square) or decreased (circle) after treatment with the chemotherapeutic agent. Again by fitting an SBM with the model selection criterion in \cite{dudin}, 5 communities were identified. With our attributed SBM, 9 communities were identified. Note that using the attributed SBM created more communities in that it split up the purple core community under the classic SBM into more small communities. The implications of this new partition are explored with an entropy calculation based on the biological classifications of the protein in Figure \ref{entropyFig}.

Using the partition of the nodes under the classic and attributed stochastic block models, we sought to use the two different classifications of the nodes (6 class modification type and 2 class increase/decrease) to compute entropy of labels within communities. The expectation is that by incorporating attribute information that is related to the functional protein information into the community detection problem, we should see a decrease in the entropy over the classification labels in communities. In Figure~\ref{entropyFig}A-B, we plot the entropy for the 2 class and 6 class node classifications, respectively. We define ${\bf E}_{c}$, the entropy for community $c$ as

\begin{equation}
{\bf E}_{c}=-\sum_{k} p_{k}\log(p_{k}).
\end{equation}

Here, $k$ indexes the unique classifications found in community $c$ and $p_{k}$ is the probability that a node in community $c$ belonged to classification $k$ in community $c$. In these plots the black and purple curves correspond to the fits of the classic and attribute SBM fits, respectively. Using both types of node classifications to compute these entropy quantities, we see that the attribute SBM succeeds in breaking up one high entropy community (5) from the classic SBM partition into lower entropy communities. 

\begin{figure}[h!]
\begin{center}
\includegraphics[width=0.4\textwidth]{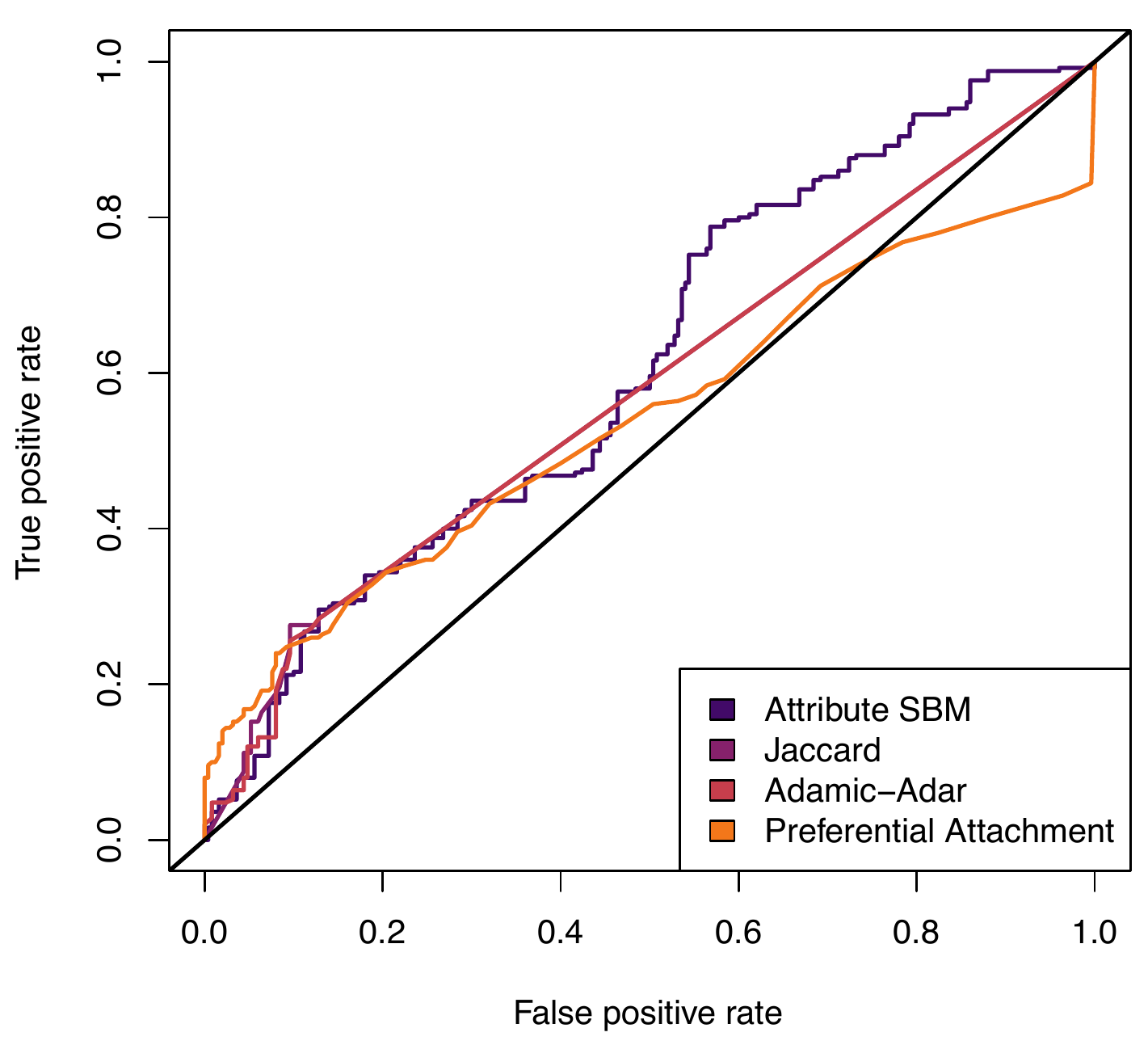}
\caption{{\bf Link Prediction in the protein interaction network}. Performing link prediction using the attributed SBM, Jaccard, Adamic Adar, and preferential attachment. The corresponding AUC curves for these methods were 0.61, 0.58, 0.58, and 0.51, respectively.}
\label{Fig9}
\end{center}
\end{figure}

{\bf Link Prediction in the Protein interaction network}
We performed link prediction on the protein interaction network using the procedure described in section 4.1. Given that this protein network is sparse, none of the link prediction methods performed particularly well. The AUC values for the attributed SBM, Jaccard, Adamic-Adar and preferential attachment were 0.61, 0.58, 0.58, and 0.54, respectively. The associated ROC curves are shown in Figure \ref{Fig9}.  

\begin{figure}[h!]
\begin{center}
\includegraphics[width=0.3\textwidth]{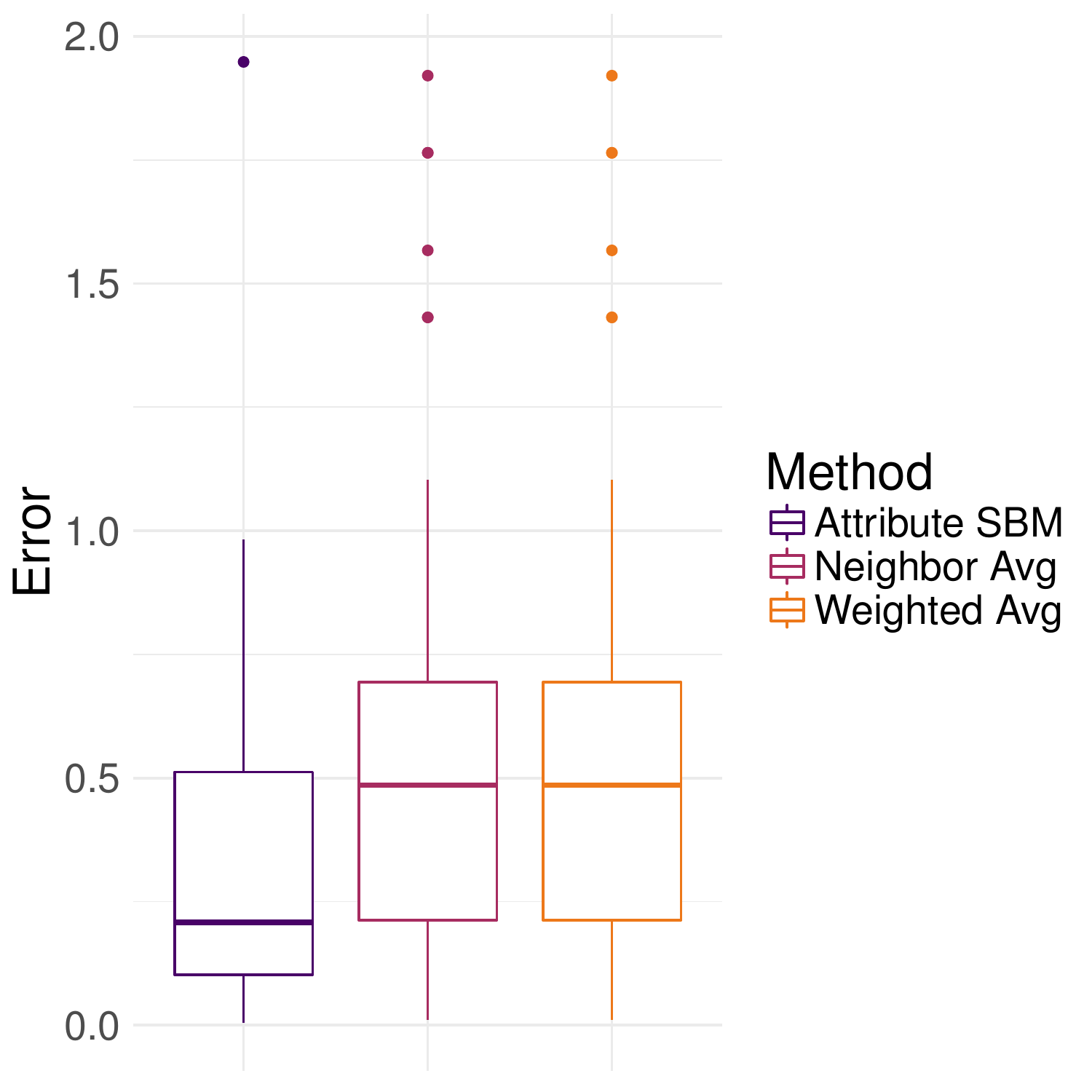}
\caption{{\bf Collaborative filtering in the protein interaction network}. For each of the 82 nodes, we fit a model to the remaining 81 node network and given the node's closest  neighbors (based on network connectivity) sought to predict its 6-dimensional attribute vector. The reported error is the relative error $\mathcal{E}$ between the difference between the true attribute vector (${\bf x}_{i}$) and its predicted attribute vector (${\hat{\bf x}}_{i}$). The mean error in ${\bf x}_{i}$ using the attributed SBM is 0.21, as opposed to the neighbor average error where it is 0.48. }
\label{collabprotein}
\end{center}
\end{figure}

{\bf Collaborative filtering in the protein interaction network}
Collaborative filtering was performed using the method described in section 4.2. Note that unlike the microbiome sample similarity network, the edges in this network are unweighted and hence the neighbor average and weighted neighbor average methods produce the same result. We note that performing collaborative filtering with the attributed stochastic block model results in a lower mean error of 0.21 compared to that of 0.48 when using the neighbor average. Similar to Figure \ref{Fig6}, the box plots in Figure \ref{collabprotein} represent the distribution of errors across each of the 82 nodes.

\section{Conclusion and future work}
We defined an attributed stochastic block model, where a node's community assignment determines its connectivity and its attribute vector. Our model builds on previous work with attributed stochastic block models because it can handle multiple continuous attributes. The continuous attributes are modeled by a Gaussian mixture model, with the assumption that the attributes for members for each community are parameterized by a unique multivariate Gaussian. Since community detection results are often difficult to validate due to the absence of a known ground truth, we quantified the ability of the fitted attributed stochastic block model to represent a particular network by performing link prediction and collaborative filtering tasks. Applying link prediction and collaborative filtering to two biological networks, we observed that the attributed SBM is useful for these applications.

Future work could extend the model to handle a combination of multiple discrete and continuous attributes. Further, while the inference or understanding of fitting a stochastic block model to weighted networks is not well understood, figuring out how to integrate edge weights and attributes in determining community structure could be useful. Finally, we briefly discussed observed detectability properties in Figure \ref{Fig3}, noting that it would be interesting to characterize how the properties of the attributes and connectivity relate to effective identification of community structure.

Networks used across fields are becoming increasingly complex, often with multiple sources of information to integrate in order to make a conclusion for the data. Our approach to an attributed SBM advances the understanding of how to jointly consider attribute and connectivity information in a probabilistic framework.  

\section*{Acknowledgment}
This work was supported by the National Science Foundation under award \#1610762.
\bibliographystyle{IEEEtran}
\bibliography{sigproc}

\end{document}